# On Measure Problems in Allometric Analysis of Cities

## -- How to correctly understand the law of allometric growth

Yanguang Chen

(Department of Geography, College of Urban and Environmental Sciences, Peking University, 100871, Beijing, China. Email: chenyg@pku.edu.cn)

**Abstract**: The law of allometric growth originated from biology has been widely used in urban research for a long time. Some conditional research conclusions based on biological phenomena have been erroneously transmitted in the field of urban geography, leading to some misunderstandings. One of the misunderstandings is that allometric analysis must be based on average measure. The aim of this paper is at explaining how to correctly understand the law of urban allometric growth by means of the methods of literature analysis and mathematical analysis. The results show the average measures cannot be applied to all types of allometric relationships, and the allometric relationships based on average measures cannot be derived from a general principle. Whether it is an empirical model or a theoretical model of allometric growth, its generation and derivation are independent of the average measures. Conclusions can be reached that the essence of allometric growth lies in that the ratio of two related general relative growth rates is a constant, and this constant represents the allometric scaling exponent and fractal dimension ratio. The average measures are helpful to estimate the allometric scaling exponent value which accords with certain theoretical expectations more effectively.

# 1. Introduction

Allometric growth was originally a biological law, but it has been widely used in urban research for a long time. Until today, the research results of allometric growth in the field of biology still have a profound impact on urban allometric analysis. A lot of the impact is positive. But there are also negative effects and misunderstandings. All along, biologists have been trying to reconcile the dynamic proportional relations in the process of allometric growth with geometric measure relations. They have been trying to calculate the allometric scaling exponent in line with the theoretical expectation from the observation data. Some scholars have found that the allometric scaling exponent estimated based on the average measures is more in line with the theoretical expectation than the scaling exponent obtained based on the dynamic measures. Bruce J. West (2017, page 29) said, "Focusing on the statistical nature of allometry, we emphasize that ARs ought to be strictly a relation between average quantities, not a relation between the dynamic variables." There is a premise to these sentences, that is, the perspective of statistical analysis. The key lies in focusing on the statistical nature of allometry. Unfortunately, this type of statements misleads some young scholars. Now there is a view in urban geography that allometric growth analysis must be based on average measures rather than dynamic variables. In fact, if a researcher read a lot of literature about allometric growth, and understand the mathematical essence of the law of allometric growth, he will find that the average measure has applicable conditions. Average measures cannot be used for all types of allometric relationships. At different scales or in different fields, the effect of average measures for allometric analysis is also different. Some conclusions of allometric growth research in the field of biology should not be blindly generalized to urban studies.

Theory is always connected with the phenomena in the real world through measurement. Divergences of understanding at the measurement level at the measurement level will affect the development of allometric growth theory and analysis methods. This paper is devoted to discussing how to correctly understand the law of allometric growth of cities. Historical literature analysis and mathematical analysis are the main methods of this paper. If the average measures in allometric relationships are absolutized, the following facts cannot be explained. First, *the origin of the* *allometric growth law*. From the perspective of literature based history and development, the law of allometric growth is not found through average measurement. Biologists and urbanists have revealed the law of allometric growth without using average measures. Second, *the mathematical principle of the allometric growth law*. From the principle of general system theory, the allometric growth model can be derived a priori by mathematical methods, but the mathematical model of allometric growth based on average measures cannot be derived from general principle. Third, *application scope of allometric growth law*. Average measure is not applicable to all types of allometric relationships. For example, cross-sectional allometric relation can be derived from a pair of Zipf's laws, but Zipf's distribution bears no characteristic scale and average value have no meaning. If the average value must be calculated, the two variables become two numbers, and thus it is impossible to estimate the allometric growth parameters. It can be seen that the average measure must have applicable conditions and ranges in allometric analysis. An indisputable fact is that the allometric scaling exponent values that meets the theoretical expectation can be estimated more effectively with the aid of the average measures. This shows that the viewpoints that the allometric relationships are the relations between average measures rather than dynamic measures is based on parameter or statistics rather based on models or theory. If a concept involves a whole set of knowledge system, it cannot be understood by reading the literature out of context. In the following sections, the questions will be discussed step by step.

## 2. The models of allometry

### 2.1 Allometric research from biology to urban geography

The law of allometric growth originated from biology, and it was developed in urban studies. The term "allometry," was coined by Huxley and Teissier (1936, pp. 780-781) from the Greek allometron (literally meaning "other-measure"). The concept was based on biological size and structure change research (Bon, 1973; Dutton, 1973). Today, the concept of allometric growth can be understood from at least four aspects: relative growth, dynamic proportion (nonlinear proportion change), invariance in change, and dimension consistency principle. (1) Relative growth. The allometric formula appears as a power function, which is used to describe "differential" or "relative" growth. As Dutton (1973) said, "…the term 'allometry' is nearly always synonymous with the term 'relative growth.' As such, the allometric perspective is perhaps the most direct way to describe how configurations in a large system differ from those of a small system of the same type." (2) Dynamic proportion relationships. The differences between configurations of a large system and that of a small system suggests changing proportions. That is to say, the system does not develop in a linear proportional relationship. Dutton (1973) pointed out, "As Galileo clearly understood six hundred years ago, no object can increase greatly in size without changes in some of its proportions (Note: clerical error, "six" should be changed to "three")." (3) Invariance in change. The ratio of the relative growth rate of one part to that of another part or the whole is unchanged (Bertalanffy, 1968). In short, the allometric exponent is a constant. It should be noted that the invariance in change of a system over time of longitudinal allometry is replaced by the invariance under transformation of mathematics of transversal allometry. (4) Dimensional consistency. The allometric exponent involves the dimension relation of different measures or different elements. This results what is called square-cube law, i.e., the 2/3 law, which is based on the ideas from Euclidean geometry. Based on fractal geometry, the 2/3 law was replaced by 3/4 law (Geoffrey West, 2017).

Allometric growth can be traced back to the 19th century. Knowing the contributions of some key scientists is helpful to understand the essence of allometric growth law (Table 1). Julian Huxley may have been inspired by Snell (1892). In 1891, Otto Snell discovered that the growth rate of mammalian brain size is different from the growth rate of body size (Lee, 1981). However, the growth of various parts of an organism maintains a dynamic proportional relationship. Further investigation revealed that there are similar patterns in human body development (Naroll and Bertalanffy, 1956/1973). For example, the human brain and body exhibit an allometric growth relationship during development. The relative growth rate of the body is higher than that of the brain. Bertalanffy (1968) derived the allometric growth relationship a priori from general systems theory. This means that allometric growth relationships are a common phenomenon in general systems including cities and systems of cities. Stephen Jay Gould is recognized as an outstanding scholar in allometric growth research. Because he knows many European languages, he can understand the papers on allometric growth written by scholars from non English speaking countries in their own languages, so as to integrate many piecemeal research results (Gayon, 2000). Moreover, Gould (1973) may be the first scholar to study the allometric scaling relationship of artificial buildings. Allometric analysis was introduced into urban study by Naroll and Bertalanffy (1956), who employed dynamic measures to research urbanization process. In the 1950s to 1970s, urban scientists conducted extensive research on allometric growth (Beckmann, 1958; Lo and Welch, 1977; Naroll and Bertalanffy, 1956). In 1973, the Journal of *Ekistics* organized a special issue entitled "*Size and shape in the growth of human communities*" on urban allometric growth (special issue in *Ekistics*, 1973). This batch of papers involved different measurements, perspectives, and types of allometric growth research on cities. Unfortunately, due to the dimensional difficult problem of the allometric growth exponent, the enthusiasm of urban scientists for allometric growth cooled down for a while (Chen, 2013). Whether in the field of biology or geography, the calculation results of the allometric scaling exponent cannot be reasonably explained mathematically using Euclidean geometry (Gould, 1973; Lee, 1989). This is the so-called dimensional dilemmas in allometric growth research (Chen, 2013; Lee, 1989). The emergence of fractal geometry has overcome the problem of dimensionality behind the allometric growth law (Batty and Longley, 1994; Chen, 2013; Chen and Xu, 1999; West, 1997).

**Table 1 Contribution of key scientists to allometric growth research**

| Scientist | Academic contribution |
| --- | --- |
| **Otto Snell** | He found that there is a dynamic proportional relationship in the process of biological growth |
| **Julian Huxley** | The concept of allometric growth is proposed and its mathematical expression is given |
| **Ludwig von Bertalanffy** | Starting from the general system theory, he derived the allometric growth relationship a priori, thus proving the universality of allometric scaling relation. This study lays a theoretical foundation for introducing allometric growth law into urban research |

| Stephen Jay Gould | He knows many European languages, thus integrating the research results of scientists from different countries. |
| --- | --- |
| Geoffrey Brian West | Using the ideas from fractal theory, he derived the 3/4 instead of the 2/3 law |

## 2.2 From 2/3 law to 3/4 law

As indicated above, the concept of allometric growth can be understood from at least four aspects: relative growth, dynamic proportion (nonlinear proportion change), invariance in change, and dimension consistency principle. Long ago, biologists found the following facts: different parts of animals develop at different speeds, but there is a dynamic proportional relationship between parts and between parts and the whole. Beckmann (1958) once pointed out: "Biologists have discovered a law of alloemtric growth in the development of organisms and have suggested that it applies to social phenomena as well. The principle of allometric growth states that the rate of relative growth of an organ is a constant fraction of the rate of relative growth of the organism." (page 247). Assuming that the relative growth rate of one part is $\alpha = \mathrm{d}x/(x\mathrm{d}t)$, and the relative growth rate of the other part or the whole is $\beta = \mathrm{d}y/(y\mathrm{d}t)$, the two relative growth rates remain unchanged, that is

$$\frac{\beta}{\alpha} = \frac{\mathrm{d}y/(y\,\mathrm{d}t)}{\mathrm{d}x/(x\,\mathrm{d}t)} = \frac{\mathrm{d}y/y}{\mathrm{d}x/x} = b \, , \tag{1}$$

where the constant $b = \beta/\alpha$ is the ratio of the relative growth rate of $y$ to that of $x$. In literature, equation (1) often appeared in different forms, for example,

$$\frac{1}{\alpha}\frac{\mathrm{d}y}{y} = \frac{1}{\beta}\frac{\mathrm{d}x}{x} \, , \tag{2}$$

or

$$\frac{\mathrm{d}y}{y\mathrm{d}t} = b\frac{\mathrm{d}x}{x\mathrm{d}t} \, , \tag{3}$$

The solution to equation (3) is a power law as below

$$y = ax^{b} \, , \tag{4}$$

where *a* denotes a proportionality coefficient, and *b* refers to a scaling exponent. Bon (1973) pointed out, "the power equation describing 'differential' or "relative" growth is known in biological sciences as the 'allometric formula.'"

A power law represents a proportional relation, which must comply with the principle of dimensional consistency. As early as ancient Greece, mathematicians discovered the principle of dimensional consistency, which states that measures of different dimensions such as length *L*, area *A*, and volume *V* cannot form proportional relationships. To form a proportional relationship between measures of different dimensions, their dimensions must be transformed into consistency (Feder, 1988; Could, 1973; Lee, 1989; Mandelbrot, 1982; Nordbeck, 1971; Takayasu, 1990), that is

$$L^{1/1} \propto A^{1/2} \propto V^{1/3}, \tag{5}$$

where $\propto$ refers to "be proportional to". Equation (5) can be equivalently expressed as

$$L^{1/1} = k'A^{1/2} = k''V^{1/3}, \tag{6}$$

where $k'$ and $k''$ are proportionality coefficients. Extending equation (5) to general, we have

$$L^{1/1} \propto A^{1/2} \propto V^{1/3} \propto M^{1/D}, \tag{7}$$

where *M* denotes a generalized volume, the dimension of which is *D*. If *D* is not an integer or a ratio of integers, it indicates that *M* is a fractal object (Feder, 1988; Mandelbrot, 1982; Takayasu, 1990).

The dimension principle is applicable to all geometric shapes. Consider two measures *x* and *y*, whose dimensions are $D_x$ and $D_y$, respectively. If they form a proportional relationship, then there must be

$$y^{1/D_y} \propto x^{1/D_x}. \tag{8}$$

Thus we have

$$y \propto x^{D_y/D_x} \propto x^b. \tag{9}$$

So the scaling exponent *b* must be the ratio of two dimensions, that is

$$b = \frac{D_y}{D_x}. \tag{10}$$

This suggests that the allometric exponent *b* may not be arbitrary, but has geometric meaning. Thus, the concept of relative growth rate changed to the concept of geometric measures.

The geometric measure relationship means that the scaling exponent have specialized value. Suppose $x$ represents the volume or mass of mammals, and $y$ represents the surface area of mammals. In this case, we have $D_x = 3$, $D_y = 2$. Therefore, the expected value of $b$ is 2/3. This is the so-called square-cube law. Haire (1973) said, "The square-cube law describes an invariable relationship between surface and mass of physical bodies. .... The square-cube law says that mass grows by a cube function while surface grows by a square." (page 265) The square-cube law was introduced into urban geography by Nordbeck (1971) to describe the allometric relation between urban population $P$ and urban area $A$. Suppose the dimension of $P$ is 3, and the dimension of $A$ is 2, the allometric exponent is expected to equal 2/3 (Lee, 1989; Nordbeck, 1971).

Unfortunately, a large number of empirical analyses in the fields of biology and urban geography do not support 2/3. Where urban allometry is concerned, in many cases, the calculated results of the exponent $b$ deviate significantly by 2/3 (Lee, 1989). So scientists thought of using average measures instead of dynamic measures. However, the estimation of allometric parameters is still biased. To solve this paroblem, Lee (1989) divided allometric relations into three types: geometric similar model, elastic similar model, and dynamic similar model. The square-cube law is merged into the dynamic similarity model. The problem returned to equation (3).

For biology, this law is related to metabolic analysis. If the basic parameters are not estimated accurately, the mechanism of metabolism cannot be understood. This is the dimensional problem of allometric growth. In 1990s, Geoffrey Brian West *et al* (1997), a theoretical physicist, used the theory of fractal geometry to derive the allometric scaling in biology, Chen and Xu (1999) used to the ideas from fractal geometry to solve the dimensional problem in urban geography. By utilizing fractal dimension, we can interpret the the mathematical essense of the allometric scaling exponent. So, research on this topic rised again. However, the problem is not over. The 3/4 law is not often verified. Comparatively speaking, based on the average measures, the calculated parameters are closer to 3/4. Anyway, as Chen (2012) pointed out, "The rules followed by the cities' distribution are all based on the notion of statistical averages, and the regularity of urban development is often perceived only on a large scale." Under certain conditions, based on the average measures, we can get better verification effect for allometric relations. But the application of average measures to allometric analysis is conditional. It will be demonstrated in detail later.

In short, the law of allometric growth is related to the principle of dimensional consistency. According to the principle of dimensional consistency, the value of allometric exponent is a priori. For biology, using average measures to estimate the model parameters of allometric growth, the results may be closer to the theoretical expectation. However, the theoretically expected value of the urban allometric scaling exponent has not been determined so far. Geoffrey Brian West once served as the Chief Scientist of Santa Fe Institute (SFI). His ideas and analytical methods influenced several young urban geographers in Europe. Since then, the study of allometric growth has once again emerged in urban geography. During this process, several excellent articles by L.M.A. Bettencourt (2007, 2013) had a great influence.

## 3. Dynamic measure and average measure

### 3.1 Group mean and individual measurement mean

Average measures include at least two types: group mean and individual measurement mean. The so-called group averaging refers to taking the average of the measurement values of multiple individuals. The so-called individual measurement averaging refers to repeatedly measuring the same individual and then taking the average. All quantitative or statistical analyses require the use of individual measurement averages. However, for urban research, individual measurement averages are often beyond the author's control. A group is a set of individuals.

In some cases, it is best to use group averages, otherwise the data may not effectively reflect patterns (or regularities) or the calculated parameters may deviate from expectations. For example, imagine us studying the general allometric growth relationship between head size and body size during human development. If we study a single person's allometry, the time is very long. To achieve this, it is necessary to continuously track and measure a person's developmental process from infancy to adulthood for no less than 20 years. Due to human body differences, even if we persist for twenty years, the estimated parameter values may not meet expectations. In this case, using the average of multiple individuals is a wise choice. Imagine dividing human development into *n*=20 age stages. Then, choose *m*=1000 people for each stage. We need to investigate 20000 individuals of different ages for this purpose (Table 2). Take the average of brain size and body size for each stage, and then establish an empirical allometric growth relationship. This method provides relatively stable results.

However, the analysis of allometric growth based on group averages has limitations. For example, Naroll and Bertalanffy (1956) once fitted the law of allometric growth to the relationship between urban population and rural population of USA from 1790 to 1950. The urban and rural population data for each time period are unique. It is impossible to calculate the average values.

**Table 2 Group means of human brain size and human body size in different age stages**

| Age group | Individual 1 | Individual 2 | …… | Individual *m* | E(*x*) | E(*y*) |
|---|---|---|---|---|---|---|
| **Age stage 1** | $x_{11}, y_{11}$ | $x_{12}, y_{12}$ | …… | $x_{1m}, y_{1m}$ | E($x_1$) | E($y_1$) |
| **Age stage 2** | $x_{21}, y_{21}$ | $x_{22}, y_{22}$ | …… | $x_{2m}, y_{2m}$ | E($x_2$) | E($y_2$) |
| **Age stage 3** | $x_{31}, y_{31}$ | $x_{32}, y_{32}$ | …… | $x_{3m}, y_{3m}$ | E($x_3$) | E($y_3$) |
| **……** | …… | …… | …… | …… | …… | …… |
| **Age stage *n*** | $x_{n1}, y_{n1}$ | $x_{n2}, y_{n2}$ | …… | $x_{nm}, y_{nm}$ | E($x_n$) | E($y_n$) |

## 3.2 Scale and mean

The effectiveness of average measurement depends on the scale of elements. For the analysis of allometric growth between small-scale elements (e.g., organism), averaging is not only effective but sometimes necessary, but for large-scale elements (e.g., city size), the averaging effect is generally not significant and even cannot be processed. If our research objective is individual allometric growth and we want to compare the differences of allometric scaling exponents among different individuals, it is not possible to use group averaging. The group averageing process should be substituted with the individal measurement averaging process. German mathematician Carl Friedrich Gauss (1777-1855) pointed out that no physical quantity is absolutely accurate, and even under the same operating conditions or similar methods, there may be slight differences in the measurement results of the same quantity by different observers. Imagine we study the developmental process of a mammal and regularly measure its weight and surface area. Different measurements of the same object taken by the same observer at the same time may yield the slightly different results. The solution to this problem is to repeatedly measure multiple times during a certain period of time and then take the average value (Table 3). For scholars who have been conducting quantitative research or statistical analysis for a long time, this is common knowledge and does not need to be explained further.

The purpose of taking the average is to minimize the impact of measurement errors on the calculation results as much as possible. If there is no measurement error, taking the average is meaningless. However, the strength of the impact of measurement errors depends on the scale of the object being measured. The smaller the scale, the greater the effect of measurement error; conversely, the larger the scale, the smaller the effect of measurement error. In short, the negative impact of measurement errors is inversely proportional to the scale of the measured object. We once attempted to validate Clark's negative exponential distribution model of urban population density decay using census data. However, we are unsure of the exact center of the city. So, we defined four possible centers and processed four sets of data. It goes without saying that there are differences among these four sets of data. Our original intention was to determine the most likely center location by comparing the calculation results based on the positioning of different possible centers of the city. However, the results were unexpected as there was no significant difference in the goodness of fit and parameter estimation results based on different datasets, and thus there was no significant difference in the analysis conclusions. We have tested several cities and the results and conclusions are all like this. This suggests that the measurement of human body is very different from urban measurement in the study of allometric growth. Because the scale of the city is not on the same order of magnitude as the scale of the human body.

**Table 3 Individual measurement means of human brain size and human body size in different age stages**

| Period | Measurement 1 | Measurement 2 | …… | Measurement $m$ | E($x$) | E($y$) |
|--------|---------------|---------------|-----|-----------------|--------|--------|

| **Time 1** | $x_{11}, y_{11}$ | $x_{12}, y_{12}$ | ...... | $x_{1m}, y_{1m}$ | $E(x_1)$ | $E(y_1)$ |
|---|---|---|---|---|---|---|
| **Time 2** | $x_{21}, y_{21}$ | $x_{22}, y_{22}$ | ...... | $x_{2m}, y_{2m}$ | $E(x_2)$ | $E(y_2)$ |
| **Time 3** | $x_{31}, y_{31}$ | $x_{32}, y_{32}$ | ...... | $x_{3m}, y_{3m}$ | $E(x_3)$ | $E(y_3)$ |
| **......** | ...... | ...... | ...... | ...... | ...... | ...... |
| **Time $n$** | $x_{n1}, y_{n1}$ | $x_{n2}, y_{n2}$ | ...... | $x_{nm}, y_{nm}$ | $E(x_n)$ | $E(y_n)$ |

There are many classic research cases in the literature. Let's take a look at the study on the allometric growth of urban and rural populations by Naroll and Bertalanffy (1956). Their observation scale is at the regional level, which belongs to the extra large scale in terms of geographic space. Moreover, census work cannot be completed by individuals, so census data cannot be extracted or generated by individuals. Researchers can only use data publicly released by the US census structure. Therefore, it is neither possible nor necessary for them to adopt group mean or individual measurement mean values for their allometric growth analysis.

## 3.3 Longitudinal allometry and transversal allometry

Allometric growth can be divided into many types, and the average measure is only applicable to some types. Whether in the field of biology or urban geography, allometric growth falls into at least two categories: longitudinal allometry and transversal allometry (e.g., Gayon, 2000; Pumain and Moriconi-Ebrard, 1997). The latter is sometimes termed cross-sectional allometry (Table 4). This involves the ergodic axiom that early quantitative geographers were fond of talking about (Harvey, 1968). The so-called ergodic axiom states that for random processes, the *time average* is equal to the *spatial average* and equal to the *phase average*. For a set of cities in a region, small towns represent cities in childhood, medium-sized cities represent cities in youth, and large cities represent cities in adulthood. So cities of different sizes can reflect the different stages of urban development (Batty and Longley, 1994). Based on this idea, the longitudinal allometry was generalized to transversal allometry. For two geometrically correlated measures, *x* and *y*, the longitudinal allometric relation can be expressed as

$$y_t = a x_t^b,$$
(11)

where is $a$ is a proportionality coefficient, $b$ is a scaling exponent, and $t$ denotes time. Correspondingly, the transversal allometric relation can be expressed as

$$y_k = a x_k^b, \tag{12}$$

where $k$=1, 2,…, refers to the rank of $x$, and $y_k$ corresponds to $x_k$. The above two models, equation (11) and (12), have a large number of applications in literature. The transversal allometric model can be derived from Zipf's law, i.e., the rank-size law of cities (Chen, 2002a).

Applying the averaging measures to the transversal allometric relation have no meaning. On the one hand, after averaging, the two variables become two numbers, and there is no way to estimate the value of allometric growth parameters. On the other hand, Zipf distribution has no characteristic scale, so the average value is meaningless (Batty and Longley, 1994; Mandelbrot, 1982; Takayasu, 1990). There is a method that can be used to introduce the average measure into allometric analysis, that is, hierarchical averaging. Based on hierarchies of cities, the cross-sectional allometry can be equivalently transformed into hierarchical allometry as below:

$$y_m = a x_m^b, \tag{13}$$

where $m$=1, 2,…, refers to the level or class of $x$ in a hierarchy (Chen, 2012). The hierarchical measure $x_m$ differs from the rank measure $x_k$. For cities, rank measurement cannot be averaged. However, the level measure represents the group average results. The processing method based on equation (13) is different from the processing method based on equation (12), but the scaling exponent is theoretically equivalent to one another and there is no significant difference in empirical analyses (Chen, 2010; Chen, 2012). This proves from another angle of view that introducing the average measure into the transversal allometric analysis has no practical significance.

In short, group averaging may be applicable for certain longitudinal allometric growth analyses. For transversal allometric relationships, hierarhical allometric scaling can be used to replace averaging processing. The rank average has one characteristic, that is, the larger the scale (city size), the fewer individuals (cities) participating in the average. On the contrary, the smaller the scale, the more individuals participating in the average.

**Table 4 Early classification of allometric growth**

| Tape | Object described | Dataset |
|---|---|---|
| **Longitudinal allometry** | Dynamic process | Time series |
| **Cross-sectional allometry (transversal allometry)** | Hierararchial structure | Cross-sectional data |

## 3.4 Inter-measure allometry and inter-element allometry

From another perspective, the longitudinal allometric growth can be divided into two types: allometric relation between measures and allometric relation between elements based on single measure. The former can be termed inter-measure allometry and the latter can be termed inter-element allometry. Based on the evolutionary process of the same system or element over time, the allometric relationship between different measures is inter-measure allometry. For example, the relationship between urban population and rural population in a region in different times (Naroll and Bertalanffy, 1956), the relationship between urban fraction and total population in different times (Dutton, 1973), the relationship between urban population and urban area of a city in different times (Chen and Lin, 2009), and so on, belong to longitudinal inter-measure allometry. On the other hand, the relationship between population size of city $i$ and population size of city $j$ in different time, belongs to inter-element allometry of single measure (Chen, 2017).

However, transversal allometric growth cannot be classified in this way. The transversal allometry model integrates inter-measure allometry and inter-element allometry into a framework (see eg., Lee, 1989; Lo and Welch, 1977; Nordbeck, 1971). The relation based on mean measures, that is

$$\mathrm{E}(y) = a\mathrm{E}(x)^b, \tag{14}$$

is only valid for the longitudinal inter-measure allometry. It is easy to prove that, for the longitudinal inter-element allometry, equation (14) is meaningless. Suppose $x=Q_j$, $y=Q_i$. Then we have

$$\mathrm{E}(y) = \mathrm{E}(Q_i) = \frac{1}{n}\sum_{i=1}^{n}Q_i = \bar{Q}, \quad \mathrm{E}(x) = \mathrm{E}(Q_j) = \frac{1}{n}\sum_{j=1}^{n}Q_j = \bar{Q}. \tag{15}$$

This suggests that E(*y*)=E(*x*). The average result provides a unique variable. This implies that group averaging is not applicable to the longitudinal allometric growth relationship between elements. Moreover, equation (14) cannot be applied to the transversal allometry. For equation (12), we have

$$\mathrm{E}(y) = \mathrm{E}(y_k) = \frac{1}{n}\sum_{n=1}^{n} y_i = \overline{y}, \quad \mathrm{E}(x) = \mathrm{E}(x_k) = \frac{1}{n}\sum_{n=1}^{n} x_i = \overline{x}. \tag{16}$$

The average result provides only two mean values, no variabe can be made use for parameter estimation. In short, the averaging allometric relation E(*y*)=*a*E(*x*)$^b$ is is applicative to the longitudial inter-measure allometry. For analyzing the inter-element allometry based on single measure, the averagng processing is meaningless because E(*y*)=E(*x*).

## 3.5 Differences between biological allometry and urban allometry

Probably influenced by the ideas from classical physics, biologists have been pursuing the discovery and verification of constants of allometric growth. In the past, they tried to test the 2/3 law, now they try to test the 3/4 law, and so on. It is not surprising that this idea has influenced the study of urban allometry. In the past, geographers were keen on the 2/3 law, and many people tried to verify it. Today, geographers who study cities from the perspective of allometric growth are talking about the 3/4 law with great interest. Many people announced that they had verified the 3/4 law of the city development. However, so far, no one has really deduced the 3/4 law from the postulates of the city itself in pure theory. In fact, geographical system has no constant in strict sense (Harvey, 1969). The central concept of geography in the classical sense is *areal differentiation* (Hartshorne, 1959). The concept of *areal differentiation* was once replaced by the notion of *spatial organization* (Harvey, 1959). Later, the concept of *areal differentiation* returned to geography in the name of *spatial heterogeneity* (Anselin, 1996). The statistical explanation of spatial heterogeneity is spatial nonstationarity, which suggests there are different probability structures in different places. In this sense, it is better to explain the spatial heterogeneity of geographical system through the difference of allometric parameters than to find the elusive geographical constant.

The parameters of urban allometric growth reflect spatial heterogeneity. A large number of my studies show that the exponent of allometric relation between urban population and urban area is a varable quantity and can be expressed as

$$b = \frac{1+k}{2+k}, \tag{17}$$

where $k$=1,2,3,…. Different spaces and places have different $k$ values. The results of urban geographers' calculations, $b$=3/4, may just be coincidental. Average measurement is not helpful to verify the 3/4 law of urban geography. Instead of blindly tracing the research conclusions of biologists, it is better to find a new way and open up a fresh outlook (Table 5).

**Table 5 A comparison between biological allometry research and urban allometry research**

| Field | Scale | Objective | Measure | Focus |
|-------|-------|-----------|---------|-------|
| **Biology** | Small | Using scaling exponent to reflect generality or similarity | The average measures are helpful to obtain the allometric parameter values that meets the theoretical expectation | Geometric measure relation |
| **Urban geography** | Very large | Use scaling exponent to reflect uniqueness or difference | The application effect of average measure is limited | Dynamic similar relation |

## 3.6 Conundrum of allometric scaling

For the longitudinal allometric analysis based on multiple elements and multiple measures, the average measure can lead to a relatively stable scaling relationship and analysis results. But there is no evidence that the average measure can replace the function of non-average measure. What if we use two or more measures based on time series to examine the allometric growth process of multiple elements? The model of longitudinal inter-measure multiple element allometry can be expressed as

$$y_k(t) = a_k x_k(t)^{b_k}, \qquad (18)$$

where $k$ denotes the rank of a city in terms of population size. A discovery is that the time series composed of the grouped average values of multiple elements still satisfies the allometric growth relationship (Chen, 2002a), that is

$$E(y_k(t)) = E(a_k x_k(t)^{b_k}) = aE(x_k(t))^b, \qquad (19)$$

where the scaling exponent $b$ approaches the mean of the the scaling exponents $b_k$ of any pair of elements. For the time being, I am unable to derive equation (19) theoretically from equation (18). However, equation (19) is supported by observational data (Chen, 2002a).

# 4. Questions and discussion

## 4.1 Three worlds of scientific research

Good classification is helpful for making scientific research. According to Casti (1996), scientific research involves three worlds: *real world*, *mathematical world*, and *computational world* (Figure 1). Because he did not understand the connection and difference between the three worlds, young Albert Einstein was once puzzled by the mathematical method for a time. Einstein once said, "I don't believe in mathematics." The reason is as follows, "As far as the laws of mathematics refer to reality, they are not certain, and as far as they are certain, they do not refer to reality." Fortunately, when he studied general relativity, Einstein finally understood the value of mathematics and its connection with reality. He said, "If the facts don't fit the theory, change the facts." This viewpoint deeply influenced August Lösch, one of the founders of the central place theory. Lösch once said somewhere, if the reality is inconsistent with a mathematical model, it may be that the reality is wrong, not the model. From the perspective of classical physics, this view of Lösch is incomprehensible; but from the perspective of complexity theory, this view does reflect Lösch's foresight (During an academic communication, a student mentioned Lösch's view, but I couldn't find the source. For reference only).

The concept of three worlds in scientific research is useful for understanding allometric measure problem. Cities appear in the real world, the mathematical reasoning and transformation are conducted in mathematical world, and the parameter estimation is carried out in the computational world (Table 6). When we describe allometric growth, we need mathematical expressions, and mathematical reasoning and transformation if necessary. These are all done in the mathematical world. When we apply a mathematical modeling result to a real problem, we need to estimate the model parameters. To do this, we need to extract the data from the real world and select the algorithm. Data processing and algorithm application are implemented in the computing world. If the estimated value of the parameter in the computational wolrd is consistent with the theoretical expected value derived in the mathematical world, there is no problem. If there is a significant difference between the estimated value of the parameter and the theoretical expected value, it is necessary to consider what the problem is. It may be that there is a problem with the premise of reasoning in the mathematical world, that there is a problem with the data observation in the real world, or that the data processing or algorithm selection in the computational world are inappropriate.

(1) The problems in the real world. For example, both organisms and cities have individual differences. What is more, as Dutton (1973) once observed, "environmental factors may cause haphazard size increase through time." There are two ways to solve this kind of problems, one is to expand the sample size, and the other is to use the average measure instead of the individual observation values.

(2) The problems in the mathematical world. For example, suppose that the organism or city is Euclidean geometric objects. The dimension of mass of organisms or cities is 3, and the dimension of surface area or urbanized area is 2. Thus, we can derive the so-called 2/3 law, i.e., the allometric exponent is $b$=2/3.

(3) The problems in the computational world. For example, some scientists argued that the least square method is not suitable for power-law parameter estimation, and the maximum likelihood method should be used to estimate the scaling exponents (Clauset *et al*, 2009; Newman, 2005). The problem is that the validity of the maximum likelihood method is based on the joint normal distribution of variables, and the scaling law means that the complex systems do not meet the normal distribution (Chen and Feng, 2017). In fact, as a compromise, it is a good choice to use the (reduced) major axis method to estimate the allometric scaling exponents (Zhang and Yu, 2010).

**Table 6 Division of three worlds in scientific research**

| Field | Phenomena | Process | Example | Nature |
|---|---|---|---|---|
| **Real world** | Organisms or cities | Derivation, reasoning, proof | Urban growth | objective |
| **Mathematical world** | Mathematical theorem or law | Observable quantities | Theoretical models, the 2/3 law, the ¾ law | exact |
| **Computational world** | Algorithms and statistical methods | Data extraction and processing, algorithm and measure selection, parameter estimation | Dynamic measure or average measure? | Subjective in some extent |

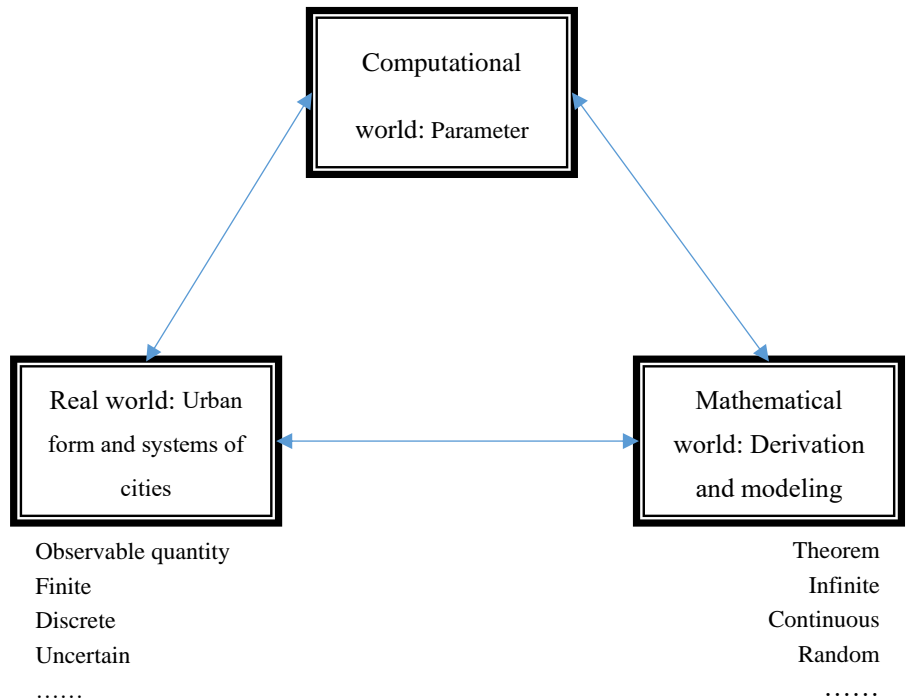

**Figure 1 Three worlds for mathematical modeling and quantitative analysis of cities**

**Note:** According to Casti (1996), scientific research involves three worlds: real world, mathematical world, and computational world. Allometric growth is the law of urban development in the real world. After abstracting this law into a mathematical model, reasoning and transformation can be carried out in the mathematical world. When we apply the model of the mathematical world to the cities in the real world and estimate the parameters, we need to process data, select algorithms and measure types in the computational world.

## 4.2 Two types of models and modeling methods

A mathematical law used to be regarded as a truth. Due to the emergence of non-Euclidean geometry and the development of quantum physics, scientists have gradually realized that laws represent models rather than truth. When the law of allometric growth is used in urban analysis, it is actually a model for understanding urban evolution. In order to understand the relationship between allometric growth law and measures, it is necessary to understand the classification of models and modeling methods. Mathematical models can be divided into two types, namely, parameter models and mechanism models. Correspondingly, there are two types of modeling methods, that is, experimental modeling method and analytical modeling method (Figure 2). These two sets of modeling methods sometimes perform their respective duties and sometimes achieve the same goal by different paths. Analytical modeling method was used to derive a priori allometric growth relationship from the principle of general system theory (Bertalanffy, 1968), or to derive the dimension ratio in allometric scaling exponent from the geometric measurement relationship (Gould, 1973; Takayasu, 1990). In contrast, experimental modeling method was used to reveal allometric relationships in observed data of organisms or cities. Where the allometric relationships betweem urban population and urban area is concerned, based on Euclidean geometry, the allometric scaling exponent value is expected to be $b$=2/3 (Lee, 1989). Based on fractal geometry, the allometric scaling exponent value comes between 2/3 and 1, that is, 2/3<$b$<1 (Chen, 2010).

However, in biology, the case is different. For example, for the allometric relationship between animal mass and surface area as an example, based on Euclidean geometry, the allometric scaling exponent value is expected to be $b$=2/3; based on fractal geometry, the allometric scaling exponent value is expected to be $b$=3/4 (West, 1997). In order to coordinate the calculation results of parametric modeling method with the theoretical expectation of analytical modeling method, average measure is favored. Before the introduction of fractal theory, there was also a contradiction between the results of parametric modeling and that of analytical modeling in urban research. After the introduction of fractal theory, this contradiction is automatically removed. With the help of fractal theory, we can derive the numerical range of the allometric scaling exponent, but can not derive the definite parameter value.

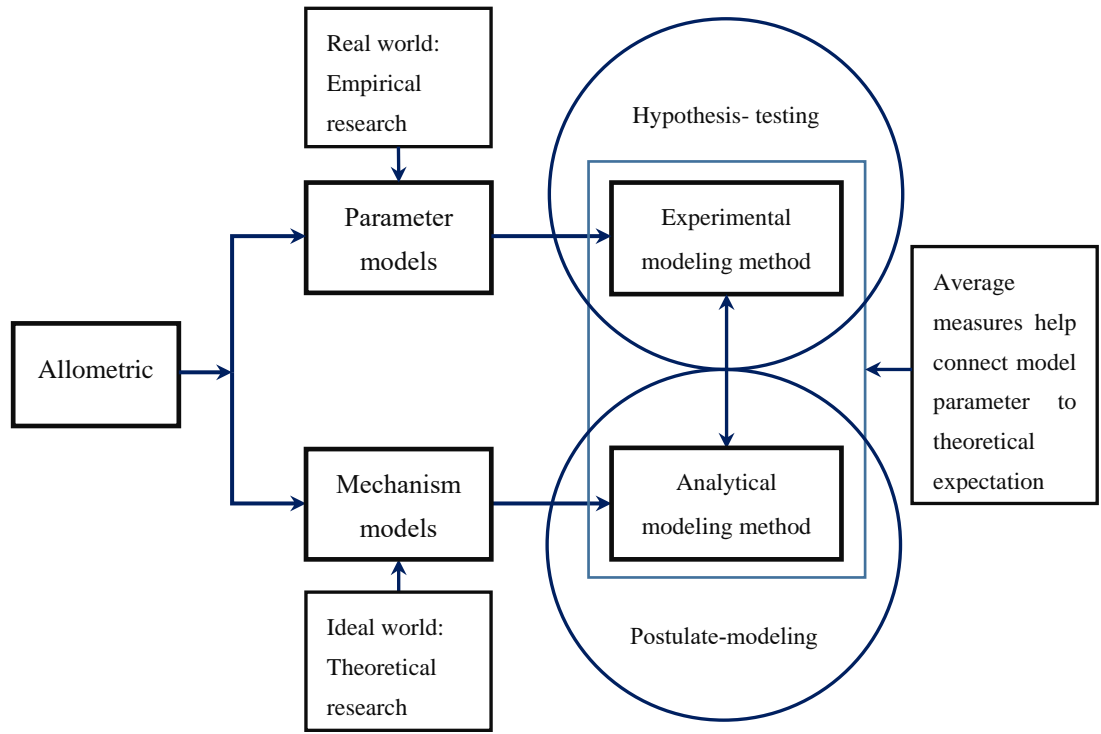

**Figure 2 Allometric analysis involve two types of mathematical models and two types of modeling methods**

**Note:** The theoretical expected value of allometric scaling exponent comes from analytical modeling method, while the actual calculated value of the allometric parameter comes from experimental modeling method. The average measures are helpful to coordinate the allometric scaling exponent values obtained by the two sets of modeling methods.

## 4.3 Many types of urban allometry studies

Nowadays, the law of allometric growth has become one of the three fundamental laws for urban studies. The other two are the distance decay law and the rank-size law (Zipf's law), respectively. The law of allometric growth has long been applied in various aspects of urban research (Table 7). The allometric growth relationship based on mean measures proposed by Bruce West (2017) cannot be applied to all types of urban allometric analyses.

The purpose of using average measurement is generally to reduce the impact of random interference, individuals differences, and measurement errors, in order to estimate stable and reliable model parameters. Unfortunately, for single measure multi-factor longitudinal allometric growth analysis, group averaging cannot provide any meaningful computational results. For transversal allometric analysis, average means are limited to special processing such as hierarchical averaging or segmental averaging. Moreover, conducting allometric growth analysis through average measure cannot effectively reflect the spatiotemporal differences in urban evolution through scaling exponents.

This paper systematically discusses the context and mathematical nature of allometric growth of cities. The essence of allometric growth law lies in that the two correlated relative growth rates are different, but the ratio of one relative growth rate to the other relative growth rate is a constant. The constant is just the allometric scaling exponent, which can be proved to equal the ratio of the dimension of one measure to the dimension of another measure. The law of allometric growth implies a dynamic proportional relationship. The system has invariance of change in the process of development. This property is reflected by the invariance under transformation in the mathematical models. One of the shortcomings of this paper is that it does not provide new empirical analysis. The previous empirical analysis is too much. It is the previous research results and findings that form the basis of the literature analysis in this paper.

**Table 7 Application examples of the law of allometric growth in urban research**

| Measure type | Data | Scale of system | Relation & Variables | Literature |
|---|---|---|---|---|
| **Dynamic measure for longitudinal allometry** | Time series | Large (region) | Urban population vs rural population | Chen *et al*, 1999; Naroll and Bertalanffy, 1956; Dutton, 1973; Shan *et al*, 1999 |
| | | | Urban fraction vs Total population | Dutton, 1973 |
| | | | Urban fraction vs mean population potential | Dutton, 1973 |
| | | | Central city size vs urban system | Beckmann, 1958; Chen, 2002b |
| | | | Total urban population vs total urban area | Chen, 2002a |
| | | Mesoscale (urban form) | Urban population vs urban area | Chen, 2002a; Chen, 2003; Chen and Lin, 2009; Shan *et al*, 1999; Liu *et al*, 1999 |

| | | | City size vs economic output | Chen, 2003; Chen and Lin, 2009; Chen and Zhou, 2003 |
|---|---|---|---|---|
| | | Small scale (Industrial organizations, e.g., firms) | external employee vs internal employee | Haire, 1973 |
| **Static measure for transversal allometry** | Cross-sectional data | Large to medium scale (urban form) | Urban population vs urban area | Dutton, 1973; Lee, 1989; Shan *et al*, 1999; Liu *et al*, 1999; Lo and Welch, 1977; Nordbeck, 1971; Shan *et al*, 1999 |
| | | | Urban population vs urban land use/economic output/ water consumption | Chen and Liu, 1998 |
| | | | Multiple urban measures (Surface area, volume, nonagricultural population, built-up area, gross domestic product, and consumption of electricity) | Lo, 2002 |
| | | Small scale (buildings) | Building area vs periphery | Gould, 1973 |
| | | | Topological structure (edges vs vertices) | Bon, 1973 |
| | | | Multiple building measures for line, area, volume (height, perimeter, area of the plot, and volume or mass),) | Batty *et al*, 2008 |

# 5. Conclusions

The law of allometric growth was originally based on biological studies. The law reflects the regularity of the development of biological morphology and structure. A large number of empirical studies show that urban development obeys the law of allometric growth in many aspects. The initial ideas of allometric growth includes relative growth, dynamic proportion and geometric measurement relationship. In recent years, scaling invariance has been revealed from the allometric relationships. The theoretical model of allometric growth does not require concrete type of measures. If we want to estimate the allometric scaling exponent that meets some theoretical expectations, the average measures are better than dynamic measures. However, the average measure is not universally applicable. The essence of allometric relation does not lie in measures. Measures are links between the mathematical model of allometric growth and real growing phenomena. The main points of this work are as follows. *First, different types of allometric analysis require different measure types*. There are many types of allometric relations, and average measures can only be applied to partial types of allometric relations. For the longitudinal allometric relationship between two elements based on single variable, the average measures cannot be adopted. *Second, the goals of allometric analyses are different in different fields*. In biological research, the allometric analysis based on average measurement can be used to calculate the special parameter values that meet the theoretical expectations (eg., 2/3, or 3/4). However, no theoretical derivation has ever proved that the scaling exponents of urban allometric growth must meet some theoretical expectations. The key of urban analysis is to reflect spatial heterogeneity, rather than looking for constant parameter values. *Third, the application effect of average measure is different with different scales*. For small-scale phenomena such as mammals, the average measure is helpful to improve the effect of allometric parameter estimation. However, for large-scale phenomena such as cities, the improvement of average measure on allometric parameter estimation is often not significant.

**Acknowledgements:**

This research was sponsored by the National Natural Science Foundation of China (Grant No. 42171192).

# Appendix

The papers in special issue entitled 'Size and shape in the growth of human communities'in Ekistics in 1973 are as follows.

- Dutton G (1973). Foreword: size and shape in the growth of human communities. *Ekistics*, 36: 142-243

- Naroll RS, Bertalanffy L von (1956). The principle of allometry in biology and social sciences. *General Systems Yearbook*, 1956,1(2): 76-89 [*Ekistics*, 1973, 36: 244-252]

- Gould SJ (1966). Allometry and size in ontogeny and phylogeny. *Biological Reviews*, 41: 587-640 [*Ekistics*, 1973, 36: 253-262]

- Haire M (1973). Biological models and empirical histories of the growth of organizations (Modern Organization Theory). *Ekistics*, 36: 263-269

- Bon R (1973). Allometry in the topologic structure of architectural spatial systems. *Ekistics*, 36: 270-276

- Doxiadis CA (1973). The structure of cities. *Ekistics*, 36: 278-281

- Woldenberg MJ (1973). An allometric analysis of urban land use in the United States. *Ekistics*, 36: 282-290

- Newling BE (1973). Urban growth and spatial structure: mathematical models and empirical evidence (The geographical Review). *Ekistics*, 36: 291-297